\documentclass{iaus}
\usepackage{graphicx}

\title[Young Stellar Populations in NGC\,922] 
{Young Stellar Populations in the \\
Collisional Ring Galaxy NGC 922}

\author[Pellerin {\it {et al.}}]   
{A. Pellerin$^1$, G. R. Meurer$^2$, K. Bekki$^3$, D. M. Elmegreen$^4$, \\ O. I. Wong$^5$, \& P. Knezek$^6$}

\affiliation{$^1$George P. and Cynthia W. Mitchell Institute for Fundamental Physics and Astronomy, Department of Physics, Texas A\&M University, College Station, TX 77843, USA \\
$^2$Physics \& Astronomy Department, Johns Hopkins University, 3400 Charles Street, Baltimore, MD 21218, USA \\
$^3$Dept. of Astrophysics \& Optics, School of Physics, University of New South Wales, NSW, 2052, Australia \\
$^4$Vassar College, Department of Physics and Astronomy, Box 745, Poughkeepsie, NY 12604, USA \\
$^5$Astronomy Department, Yale University, P.O. Box 208101, New Haven, CT 06520-8101, USA \\
$^6$WIYN Consortium Inc., PO Box 26732, Tucson, AZ 85726, USA}

\pubyear{2010}
\volume{262}  
\pagerange{TBD}
\jname{Stellar Populations: Planning for the Next Decade}
\editors{G. Bruzual \& S. Charlot, eds.}

\begin{document}

\maketitle

\begin{abstract}
We studied the star cluster population properties in the nearby collisional ring galaxy NGC 922 using HST/WFPC2 photometry and population synthesis modeling. We found that 69\% of the detected clusters are younger than 7 Myr, and that most of them are located in the ring or along the bar, consistent with the strong H$\alpha$ emission. The images also show a tidal plume pointing toward the companion. Its stellar age is consistent with pre-existing stars that were probably stripped off during the passage of the companion. We compared the star-forming complexes observed in NGC 922 with those of a distant ring galaxy from the GOODS Þeld. It indicates very similar masses and sizes, suggesting similar origins.   Finally, we found clusters that are excellent progenitor candidates for faint fuzzy clusters. 

\end{abstract}

\firstsection 
\section{Introduction}
Collisional ring galaxies are the result of a very specific case of galaxy interaction where a compact companion drops through a much more massive disk galaxy (e.g. \cite[Hernquist \& Weil 1993]{hern93}). While the companion passes through the disk, the momentary enhanced gravitational potential attracts the stars and gas towards the nucleus. This usually creates a nuclear burst of star formation. When the companion leaves, the material that was going inward is then projected outward, generating an expanding ring of star formation. Because such collision requires specific physical conditions such as the relative mass of the two bodies, and the angle and location of the collision, ring galaxies are rare in the local universe, but are thought to be more common at high redshift.

Here we present the stellar cluster population study of NGC\,922, a collisional ring galaxy at 43\,Mpc displaying a 15\,kpc C-shape ring due to an off-centered collision with a dwarf companion identified by \cite{wong06}. 

\section{HST/WFPC2 Observations and Data Processing}

We obtained images of NGC\,922 in June 2007 using HST/WFPC2 in the F300W, F439W, F547M, F814W filters. The data were processed with the HST pipeline, and drizzled to a resolution of 39\,pc. We performed PSF photometry of star clusters and found 2248 clusters seen in the four filters, and an additional 1569 clusters too faint to be detected in the F300W filter. We performed spectral synthesis on every cluster using Starburst99 (\cite[Leitherer \etal\ 1999]{lei99}) with an instant burst at solar metallicity and a standard Kroupa initial mass function. For each cluster, we estimated an age and stellar mass. The age and mass uncertainties are mainly related to the photometric uncertainties.

\section{Stellar Content in NGC\,922}

We found that 69\% of the clusters are younger than 7\,Myr, capable of ionizing the surrounding medium. Their location in the bar and ring is consistent with the H$\alpha$ emission image from \cite{wong06}. The cluster luminosity functions in NGC\,922 display slopes of 2.0 to 2.2, depending on the filter. This is consistent with the slopes found in other nearby young star forming galaxies (\cite[Whitmore \etal\ 1999]{whit99}, \cite[Larsen 2002]{lar02}).

We detected a 7-8\,kpc faint and diffuse plume in the WFPC2 images pointing toward the companion. It is undetected in the F300W filter and display no star clusters in F439W. We performed a large aperture photometry and stellar population synthesis on the plume. It indicates an age of 1-5\,Gyr and a stellar mass of about 5$\times$10$^9$\,M$_{\odot}$. The age suggests either that the plume was part of the main disk of NGC\,922 and was expelled by S2 during the collision or it was part of S2 and was ripped off during the encounter. In any case, it did not produce significant star formation related the collision. 

We studied the general properties of 14 star forming complexes observed in NGC\,922. They have diameters between 0.75 and 2.0\,kpc. They show strong H$\alpha$ emission, ages between 4 and 86\,Myr, and stellar masses from 6$\times$10$^6$ to 2$\times$10$^8$\,M$_{\odot}$. The complexes in NGC\,922 resemble those of the galaxy GDS\,21238 at z$=$0.664 (\cite[Wolf \etal\ 2003]{wolf03}) which displays 16 complexes in its 10\,kpc ring. They have ages from 80 to 150\,Myr and masses from 3 to 7$\times$10$^7$\,M$_{\odot}$.  Also, several clumpy galaxies in the HST Ultra-Deep Field show similar properties to those in NGC\,922 (\cite[Wong \etal\ 2006]{wong06}, \cite[Elmegreen \& Elmegreen 2005]{elm05}). However, the largest high-redshift complexes are about 13-20\% the size of the host galaxy, which is 2 times more than NGC\,922 and 3 times more than normal nearby spiral galaxies.

We identified several star clusters older than 50\,Myr located in the ring of NGC\,922. Because of the collisional nature of the ring in NGC\,922, those clusters were likely formed in the highly shocked low density gas regions in a low tidal force environment as presented by \cite{elm08}. As discussed by the author, such environment is propitious to the formation of faint fuzzy star clusters, i.e. faint and non-compact clusters of 1-5\,Gyr (\cite{lar00}). Because the 50\,Myr clusters in NGC\,922 seemed to have survived the infant mortality phase of clusters (see M. Gieles, IAU Symp. 266 proceeding), they are good candidate to become faint fuzzy clusters.


\begin{thebibliography}{}

\bibitem[Elmegreen (2008)]{elm08}
{{Elmegreen}, B.~G.} 2008,
\textit{ApJ}, 672, 1006

\bibitem[Elmegreen \& Elmegreen (2005)]{elm05}
{{Elmegreen}, B.~G. \& {Elmegreen}, D.~M.} 2005,
\textit{ApJ}, 627, 632

\bibitem[Hernquist \& Weil (1993)]{hern93}
{{Hernquist}, L. \& {Weil}, M.~L.} 1993
\textit{MNRAS}, 261, 804

\bibitem[Larsen \& Brodie 2000]{lar00}
{Larsen, S.~S. \& Brodie, J.~P.}
\textit{AJ}, 120, 2938

\bibitem[Larsen (2002)]{lar02}
{Larsen, S. S.} 2002,
\textit{AJ}, 124, 1393

\bibitem[Leitherer \etal\ (1999)]{lei99}
{Leitherer, C., \etal} 1999,
\textit{ApJS}, 123, 3

\bibitem[Whitmore \etal (1999)]{whit99}
{Whitmore, B.~C., Zhang, Q., Leitherer, C., Fall, S.~M., Schweizer, F., Miller, B.~W.} 1999,
textit{ApJ}, 118, 1551

\bibitem[Wolf \etal\ (2003)]{wolf03}
{Wolf, C., Meisenheimer, K., Rix, H.-W., Borch, A., Dye, S., \& Kleinheinrich, M.} 2003,
\textit{AAP}, 401, 73

\bibitem[Wong \etal\ (2006)]{wong06}
{Wong, O. I. \etal.} 2006,
\textit{MNRAS}, 370, 1607

\end{thebibliography}
\end{document}